\documentclass[aps,showpacs,twocolumn,amsfonts,amsmath,floatfix,german]{revtex4}
\usepackage[final]{graphicx}

\newlength\figurewidth
\AtBeginDocument{\setlength\figurewidth{.9\linewidth}}

\usepackage{color}
\let\omp\marginpar\relax\def\marginpar#1{\omp{\color{red}#1}}

\begin{document}
\title{Self--diffusion and Interdiffusion in Al$_{80}$Ni$_{20}$ Melts: Simulation and Experiment}
\date{\today}

\author{J. Horbach$^{1,2}$, S.K. Das$^{1,3}$, A. Griesche$^{4}$, 
M.-P. Macht$^{4}$, G. Frohberg$^{5}$, and A. Meyer$^{2}$}
\affiliation{$^{(1)}$Institut f\"ur Physik, Johannes--Gutenberg--Universit\"at Mainz,
                     Staudinger Weg 7, 55099 Mainz, Germany\\
             $^{(2)}$Institut f\"ur Materialphysik im Weltraum, Deutsches Zentrum f\"ur 
                     Luft-- und Raumfahrt, 51170 K\"oln, Germany\\
             $^{(3)}$Institute for Physical Science and Technology, University of Maryland,
                     College Park, MD 20742, USA\\
             $^{(4)}$Hahn--Meitner--Institut Berlin, Dept. Materials (SF3), 
                     Glienicker Str. 100, 14109 Berlin, Germany\\
             $^{(5)}$Institut f\"ur Werkstoffwissenschaften und --technologien,
                     Technische Universit\"at Berlin, Hardenbergstr.~36,
                     10623 Berlin, Germany}

\begin{abstract}
A combination of experimental techniques and molecular dynamics (MD)
computer simulation is used to investigate the diffusion dynamics in
Al$_{80}$Ni$_{20}$ melts.  Experimentally, the self--diffusion coefficient
of Ni is measured by the long--capillary (LC) method and by quasielastic
neutron scattering.  The LC method yields also the interdiffusion
coefficient.  Whereas the experiments were done in the normal liquid
state, the simulations provided the determination of both self--diffusion
and interdiffusion constants in the undercooled regime as well.
The simulation results show good agreement with the experimental data.
In the temperature range 3000\,K$\ge T \ge 715$\,K, the interdiffusion
coefficient is larger than the self--diffusion constants. Furthermore the
simulation shows that this difference becomes larger in the undercooled
regime. This result can be refered to a relatively strong temperature
dependence of the thermodynamic factor $\Phi$, which describes the
thermodynamic driving force for interdiffusion. The simulations also
indicate that the Darken equation is a good approximation, even in the
undercooled regime. This implies that dynamic cross correlations play
a minor role for the temperature range under consideration.
\end{abstract}

{\bf accepted for publication in Phys. Rev. B}

\maketitle

\section{Introduction}
\label{sec1}

Multicomponent liquids exhibit transport processes due to concentration
fluctuations among the different components. In the hydrodynamic
limit, these processes are described by interdiffusion coefficients
\cite{hansen,shimoji,allnatt87}.  In the simplest case of a binary AB
mixture, there is one interdiffusion coefficient $D_{\rm AB}$. This
quantity plays an important role in many phenomena seen in metallic
mixtures, such as solidification processes \cite{cahn83}, the slowing
down near the critical point of a liquid--liquid demixing transition
\cite{hohenberg77} or glassy dynamics \cite{glassbook}.

Many attempts have been undertaken for different binary
systems to relate $D_{\rm AB}$ to the self--diffusion constants
$D_{\rm A}$ and $D_{\rm B}$ via phenomenological formula (see
e.g.~\cite{vogelsang88,kehr89,hess90,trullas94,baumketner99,wax07}).
An example is the Darken equation \cite{darken49} that is widely used to
estimate the interdiffusion constant of simple binary fluid mixtures.
This equation expresses $D_{\rm AB}$ as a simple linear combination of
the self--diffusion coefficients, $D_{\rm AB}=\Phi (c_{\rm B} D_{\rm A}+
c_{\rm A} D_{\rm B})$ (with $c_{\rm A}$, $c_{\rm B}$ the mole fractions
of A and B particles, respectively). Here, the so--called thermodynamic
factor $\Phi$ contains information about static concentration fluctuations
in the limit of long wavelength.

The relationship between one--particle transport and collective
transport properties is a fundamental question in undercooled liquids
\cite{glassbook}.  In the framework of the mode--coupling theory of the
glass transition, Fuchs and Latz \cite{fuchs93} have studied a binary
50:50 mixture of soft--spheres with a size ratio of 1.2.  Their numerical
data indicate that the Darken equation is a good approximation for the
latter system in the undercooled regime.  However, from experiments
or computer simulations, not much is known about the validity of the
Darken equation for undercooled liquids.  This is due to the lack
of experimental data for interdiffusion coefficients in this case.
Moreover, most of the computer simulation studies on the relation between
self--diffusion and interdiffusion have been only devoted to the normal
liquid state. In this case, the Darken equation often seems to work
quite well \cite{hansen,trullas94,wax07,asta99,faupel03}.

In this work, a combination of experiment and molecular dynamics (MD)
simulation is used to study the diffusion dynamics in the metallic liquid
Al$_{80}$Ni$_{20}$. In the MD simulation, the interactions between the
atoms are modelled by an embedded atom potential proposed by Mishin
{\it et al.}~\cite{mishin02}.  The present work is a continuation of a
recent study \cite{das05}, where a combination of quasielastic neutron
scattering (QNS) and MD simulation was applied to investigate chemical
short--range order and self--diffusion in the system Al--Ni at different
compositions.  In the latter study, we have shown that the MD simulation
yields good agreement with the QNS data, both for structural quantities
and the Ni self--diffusion constant, $D_{\rm Ni}$. In the present work,
an additional experimental method, the long--capillary (LC) technique,
is used.  This method allows to determine simultaneously the self--diffusion
constant $D_{\rm Ni}$ and the interdiffusion coefficient $D_{\rm AB}$
(see below).

Above the liquidus temperature (i.e.~in the normal liquid state),
thermodynamic properties as well as structure and dynamics of
Al$_{80}$Ni$_{20}$ have been studied by different approaches (see, e.g.,
\cite{asta99,batalin83,ayushina69,maret90,saadeddine94,asta01}). The
Al--Ni system is an ordering system which is manifested in a negative
enthalpy of mixing \cite{enthalpy}. Thus, it does not exhibit a
liquid--liquid miscibility gap where one would expect 
that the interdiffusion coefficient vanishes when approaching the
critical point, whereas the self--diffusion constants are not affected
by the critical slowing down (see \cite{das06} and references therein).
Such a behavior is not expected for the system Al--Ni.

In the computer simulation, the Al$_{80}$Ni$_{20}$ melt can be undercooled
to an arbitrary extent avoiding the occurrence of crystallization
processes.  Therefore, we were able to study a broad temperature
range in our MD simulations, ranging from the normal liquid state at
high temperature to the undercooled liquid at low temperature. In the
experiments presented below crystallization occurs due to heterogeneous
nucleation.  Thus, the experiments were performed above the liquidus
temperature $T_{\rm L}\approx 1280$\,K. The combination of experiment and
simulation presented in this work allows for a test of the validity of
the Darken equation in Al$_{80}$Ni$_{20}$.  We will see below that this
equation is indeed a good approximation, even in the undercooled regime.

In the next section, we summarize the basic theory on self--diffusion
and interdiffusion.  The details of the experiments and simulation
are given in Sec.~\ref{sec3} and Sec.~~\ref{sec4}, respectively.
In Sec.~\ref{sec5} we present the results. Finally, we give a summary
of the results in Sec.~\ref{sec6}.

\section{Self--diffusion and interdiffusion: Basic theory}
\label{sec2}
Consider a three--dimensional, binary AB system of $N=N_{\rm A}+N_{\rm
B}$ particles (with $N_{\rm A}$, $N_{\rm B}$ the number of A and B
particles, respectively).  The self--diffusion constant $D_{{\rm s},
\alpha}$ ($\alpha={\rm A, B}$) is related to the random--walk motion of
a tagged particle of species $\alpha$ on hydrodynamic scales. It can be
calculated from the velocity autocorrelation function \cite{hansen},
\begin{equation}
  \label{eq3d1}
  C_{\alpha}(t)= \frac{1}{3 N_{\alpha}} \sum_{j=1}^{N_{\alpha}}
                 \langle 
                 {\bf v}^{(\alpha)}_{j}(t) \cdot 
                 {\bf v}^{(\alpha)}_{j}(0)
                 \rangle \; ,
\end{equation}
via a Green--Kubo integral:
\begin{equation}
  \label{eq3d2}
  D_{{\rm s}, \alpha} = \int_0^{\infty} C_{\alpha}(t) dt \; .
\end{equation}
In Eq.~(\ref{eq3d1}), ${\bf v}^{(\alpha)}_{j}(t)$ is the velocity of 
particle $j$ of species $\alpha$ at time $t$.

The self--diffusion constant can be also expressed by long--time limit
of the mean--squared displacement (MSD):
\begin{equation}
   \label{eq3d3}
  D_{{\rm s}, \alpha} = \lim_{t \to \infty} 
    \frac{1}{N_{\alpha}} \sum_{j=1}^{N_{\alpha}}
  \frac{ \left\langle \left[ 
  {\bf r}^{(\alpha)}_{j}(t) - {\bf r}^{(\alpha)}_{j}(0)
   \right]^2 \right\rangle}{6t} \; .
\end{equation}
Here, ${\bf r}^{(\alpha)}_{j}(t)$ is the position of particle $j$ of
species $\alpha$ at time $t$. Note that Eq.~(\ref{eq3d3}) is equivalent
to the Green--Kubo formula (\ref{eq3d2}).

Interdiffusion is related to the collective transport of mass driven
by concentration gradients. The transport coefficient that describes
this process is the interdiffusion constant $D_{\rm AB}$ which can be
also expressed by a Green--Kubo relation, i.e.~by a time integral over
an autocorrelation function. The relevant variable in this case is the
concentration or interdiffusion current \cite{hansen} given by
\begin{equation}
  {\bf J}_{\rm AB}(t) = \sum_{i=1}^{N_{\rm A}} {\bf v}_i^{\rm (A)}(t)
   - c_{\rm A} \left[ \sum_{i=1}^{N_{\rm A}} {\bf v}_i^{\rm (A)}(t)
                     + \sum_{i=1}^{N_{\rm B}} {\bf v}_i^{\rm (B)}(t)
               \right]
  \label{eq_curr}
\end{equation} 
where $c_{\rm A}\equiv N_{\rm A}/N = 1 - c_{\rm B}$ is the total
concentration (mole fraction) of A particles. As a matter of fact, the autocorrelation
function of the variable ${\bf J}_{\rm AB}(t)$ depends on the reference
frame and fluctuations of ${\bf J}_{\rm AB}(t)$ have to be adapted to
the ensemble under consideration. Whereas experiments are usually
done in the canonical ensemble, in a molecular dynamics simulation,
the natural ensemble is the microcanonical ensemble with zero total
momentum \cite{raineri89}.  Thus,
\begin{equation}
  \sum_{i=1}^{N_{\rm B}} {\bf v}_i^{\rm (B)} =
    - \frac{m_{\rm A}}{m_{\rm B}} \sum_{i=1}^{N_{\rm A}} {\bf v}_i^{\rm
    (A)}
  \label{eq_momcon}
\end{equation} 
follows, where $m_{\rm A}$ and $m_{\rm B}$ denote the masses of A and
B particles, respectively. Introducing the ``centre of mass velocity of
component $\alpha$ ($\alpha={\rm A, B}$)'' by
\begin{equation}
   {\bf V}_{\alpha}(t) = \frac{1}{N_{\alpha}} \sum_{i=1}^{N_{\alpha}}
              {\bf v}_i^{(\alpha)}(t) \; ,
\end{equation} 
we can use expression (\ref{eq_momcon}) to simplify the formula for the
interdiffusion current,
\begin{equation}
    \label{eq_curr2}
  {\bf J}_{\rm AB}(t) = N c_{\rm B} c_{\rm A} \left(
        1+ \frac{m_{\rm A}c_{\rm A}}{m_{\rm B}c_{\rm B}} \right)
       {\bf V}_{\rm A}(t) \; .
\end{equation} 
Thus, we have to consider only the velocities of one species to compute
${\bf J}_{\rm AB}(t)$.

Now, the autocorrelation function for the interdiffusion current is
given by
\begin{eqnarray}
  C_{\rm AB}(t) & = & \left\langle 
              {\bf J}_{\rm AB}(t) \cdot {\bf J}_{\rm AB}(0) \right\rangle \\ 
 &= &  N^2 \left( c_{\rm B} c_{\rm A} \right)^2
    \left( 1+ \frac{m_{\rm A}c_{\rm A}}{m_{\rm B}c_{\rm B}} \right)^2
\left\langle
       {\bf V}_{\rm A}(t) \cdot
       {\bf V}_{\rm A}(0) \right\rangle \; . \nonumber
\end{eqnarray}
The Green--Kubo formula for $D_{\rm AB}$ reads
\begin{equation}
   \label{eq_dab1}
   D_{\rm AB} = \frac{1}{3 N S_{cc}(0)} \int_0^{\infty} C_{\rm AB}(t) \, dt
\end{equation}
where $S_{cc}(0)$ is the concentration--concentration structure factor in the
limit $q\to 0$. The function $S_{\rm cc}(q)$ is the static correlation function
associated with concentration fluctuations. It can be expressed by a linear combination
of partial static structure factors $S_{\alpha \beta}(q)$ ($\alpha, \beta ={\rm A, B}$) 
as follows \cite{hansen}:
\begin{equation}
  \label{eq_scc}
  S_{cc}(q) = c_{\rm B}^2 S_{\rm AA}(q) + c_{\rm A}^2 S_{\rm BB}(q) 
              - 2 c_{\rm A} c_{\rm B} S_{\rm AB}(q)
\end{equation}
with
\begin{equation}
    S_{\alpha \beta}(q) = \frac{1}{N} 
              \sum_{k=1}^{N_{\alpha}} \sum_{l=1}^{N_{\beta}}
          \left\langle 
          \exp \left[ i {\bf q} \cdot ({\bf r}_k - {\bf r}_l) \right] 
          \right\rangle \ .
\end{equation}
Using elementary fluctuation theory \cite{hansen}, $S_{cc}(0)$ can be
related to the second derivative of the molar Gibbs free energy $g$,
\begin{equation}
   \label{eq3d8}
  \Phi = \frac{c_{\rm A}c_{\rm B}}{k_B T} 
         \frac{\partial^2 g}{\partial c_{\rm A} \partial c_{\rm B}} \; ,
\end{equation}
via 
\begin{equation}
   \label{eq_phi}
   \Phi = \frac{c_{\rm A} c_{\rm B}}{S_{\rm cc}(q=0)} \ .
\end{equation}
In Eq.~\ref{eq3d8}, $k_B$ is the Boltzmann constant and $T$ the
temperature. In the following, we will refer to $\Phi$ as the
thermodynamic factor.

We note that the total structure factor for the number density,
$S_{nn}(q)$, and cross correlation between number density and
concentration, $S_{nc}(q)$, can also be written as a linear combinations of
partial structure factors. These functions are given by \cite{hansen}
\begin{eqnarray}
S_{nn}(q) & = & S_{\rm AA} (q) + 2 S_{\rm AB} (q) +
                S_{\rm BB} (q) \quad , \label{eq_snn} \\
S_{nc}(q) & = &
c_{\rm B} S_{\rm AA}(q) - c_{\rm A} S_{\rm BB}(q)
+ (c_{\rm B}-c_{\rm A}) S_{\rm AB}(q)
\quad . \label{eq_snc}
\end{eqnarray}
The typical behavior of these functions for a liquid mixture will be
discussed in the result's section. The functions $S_{nn}(q)$,
$S_{nc}(q)$ and $S_{cc}(q)$ are often called Bhatia--Thornton structure
factors \cite{bhatia70}. In principle, these functions can be determined
in neutron scattering experiments, either by using isotopic enrichment techniques
(see, e.g., Ref.~\cite{maret90}) or by applying a combination of
neutron scattering and X--ray diffraction \cite{holland06}.

With Eqs.~(\ref{eq_dab1}) and (\ref{eq_phi}), the interdiffusion constant
can be written as
\begin{equation}
  D_{\rm AB} =  N \frac{c_{\rm A}c_{\rm B} \Phi}{3}
    \left( 1+ \frac{m_{\rm A}c_{\rm A}}{m_{\rm B}c_{\rm B}} \right)^2
  \int_0^{\infty}
\left\langle
       {\bf V}_{\rm A}(t) \cdot
       {\bf V}_{\rm A}(0) \right\rangle \, dt \; .
\end{equation}
Alternatively, $D_{\rm AB}$ can be also easily related to the self--diffusion constants
to yield
\begin{eqnarray}
  \label{eq_dabself}
  D_{\rm AB} & = & \Phi ( c_{\rm A} D_{\rm B} + c_{\rm B} D_{\rm A} \nonumber \\
             & + & c_{\rm A} c_{\rm B} \int_0^{\infty} 
               \left[ \Lambda_{\rm AA} + \Lambda_{\rm BB}
                     - 2 \Lambda_{\rm AB} \right] dt ) \; ,
\end{eqnarray}
where the functions $\Lambda_{\alpha \beta}(t)$ denote distinct velocity
correlation functions,
\begin{equation}
   \Lambda_{\alpha \beta}(t) = \frac{1}{3Nc_{\alpha}c_{\beta}} 
     \sum_{k=1}^{N_{\alpha}} 
     \sum_{l=1 \atop l\neq k \; {\rm if} \; 
        \alpha=\beta}^{N_{\beta}} \left\langle
     {\bf v}_k^{\rm (\alpha)}(t) \cdot {\bf v}_l^{\rm (\beta)}(0)
     \right\rangle \; .
\end{equation}
Note that the three functions $\Lambda_{\alpha \beta}(t)$ can be expressed
by the ``centre--of--mass'' correlation function $C_{\rm AB}(t)$ and the
velocity autocorrelation functions $C_{\alpha}(t)$ (the latter, multiplied
by $1/c_{\alpha}$, has to be subtracted in the case of $\Lambda_{\rm
AA}(t)$ and $\Lambda_{\rm BB}(t)$) \cite{baumketner99}. Thus, the
functions $\Lambda_{\alpha \beta}(t)$ do not contain any additional
information compared to $C_{\rm AB}(t)$ and $C_{\alpha}(t)$ and so we
do not consider them separately in the following.

If one denotes the distinct part in (\ref{eq_dabself}) by 
\begin{equation}
   \Delta_{\rm d} = c_{\rm A} c_{\rm B} \int_0^{\infty}
       \left[ \Lambda_{\rm AA}(t) + \Lambda_{\rm BB}(t)
             - 2 \Lambda_{\rm AB}(t) \right] \, dt
\end{equation}
one can rewrite Eq.~(\ref{eq_dabself}),
\begin{equation}
   \label{darken}
   D_{\rm AB} = \Phi S \left( c_{\rm A} D_{\rm B} + c_{\rm B} D_{\rm A} \right),
\end{equation}
with 
\begin{equation}
  \label{eq_manning}
  S = 1 + \frac{\Delta_{\rm d}}{c_{\rm A} D_{\rm B} + c_{\rm B} D_{\rm A}}
\end{equation}
The quantity $S$ measures the contribution of cross correlations
to $D_{\rm AB}$. If $S=1$ holds, the interdiffusion constant is
determined by a linear combination of the self--diffusion constants. In
this case, Eq.~(\ref{darken}) leads to the Darken equation
\cite{darken49}. Note that, in the context of chemical diffusion in
crystals, $S$ is called Manning factor \cite{manning61}.

As in the case of self--diffusion, the interdiffusion constant can be
also expressed via a mean--squared displacement which involves now the
centre--of--mass coordinate of species A,
\begin{equation}
   \label{eq_ra}
  {\bf R}_{\rm A}(t) = \frac{1}{N_{\rm A}} 
   \sum_{j=1}^{N_{\rm A}} {\bf r}_j^{\rm (A)}(t) \ .
\end{equation}
Then, the ``Einstein relation'' for $D_{\rm AB}$ reads
\begin{equation}
   \label{eq_dabmsd}
   D_{\rm AB} = \lim_{t\to \infty}
   \left(1+\frac{m_{\rm A}c_{\rm A}}{m_{\rm B}c_{\rm B}} \right)^2
   N c_{\rm A} c_{\rm B} \Phi
   \frac{\left\langle \left[ 
    {\bf R}_{\rm A}(t) - {\bf R}_{\rm A}(0) \right]^2
   \right\rangle}{6t} \ . 
\end{equation}
This formula can be used to determine $D_{\rm AB}$ in a computer
simulation, where the system is located in a simulation box with periodic
boundary conditions. However, in this case one has to be careful because
the difference ${\bf R}_{\rm A}(t) - {\bf R}_{\rm A}(0)$ has to be
calculated in an origin independent representation \cite{allen94}. This
can be achieved by computing this difference via the integral $\int_0^t
{\bf V}_{\rm A}(t^{\prime}) dt^{\prime}$.

\section{Experimental Methods}
\label{sec3}
\subsection{Long--capillary technique}

The long--capillary technique (LC) has been used to measure interdiffusion
and Ni self--diffusion in liquid Al$_{80}$Ni$_{20}$. The sample material
production is similar to that of Al$_{87}$Ni$_{10}$Ce$_{3}$, which
is described in Ref.~\cite{griesche06}. The experimental apparatus,
the measurement of the concentration profiles and the evaluation
of the concentration profiles, including the determination
of Fick's diffusion coefficients, are also described elsewhere
\cite{griesche04,griesche07}. Thus, here the experimental technique is
reported only briefly. In more detail we describe an improved diffusion
couple setup, which has been used in this work. This setup, with a
vertical diffusion capillary of 1.5\,mm diameter, has an increased
stabilization against natural convection and minimizes the systematic
error of convective mass flow contributions to the total mass transport.

\begin{figure}
\vspace*{0.3cm}
\includegraphics[width=\figurewidth]{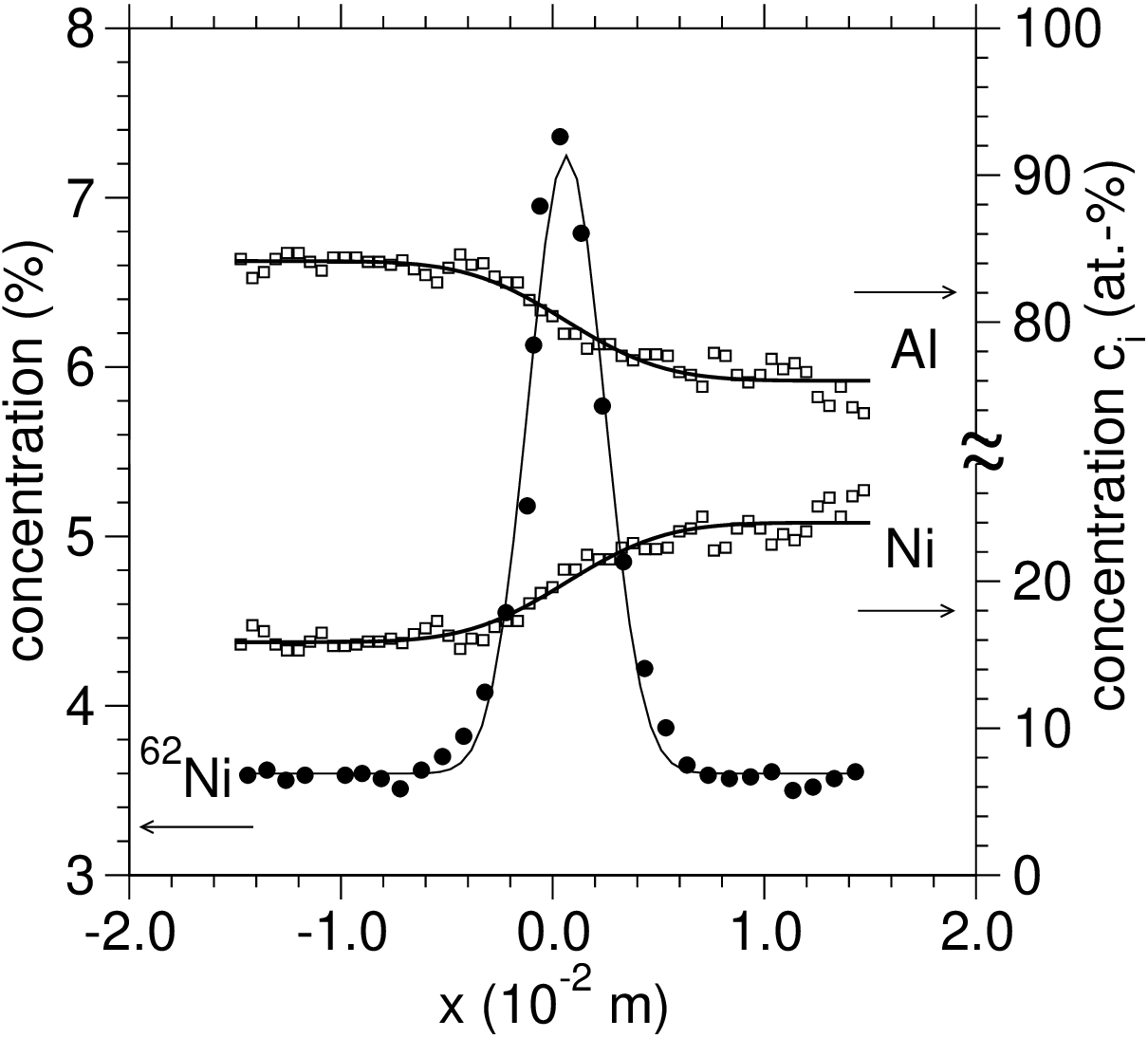}
\caption{\label{fig1}
Typical concentration profiles of a combined interdiffusion and
self--diffusion experiment. The squares denote the Al and Ni concentrations
measured by energy-dispersive X--ray spectrometry (EDS) and the dots
denote the $^{62}$Ni concentration measured by inductively--coupled
plasma mass spectrometry (ICP--MS). The lines represent the best fit
(least--square method) of the appropriate solution of Fick's diffusion
equations to the measured concentrations.
}
\end{figure}
The improvement of the diffusion couple setup implies the combination
of interdiffusion and self--diffusion measurements in one experiment. An
Al$_{80}$Ni$_{20}$ slice of 2\,mm thickness, containing the enriched
stable $^{62}$Ni isotope, is placed between both rods of an interdiffusion
couple. The interdiffusion couple consists of a 15\,mm long rod of
Al$_{85}$Ni$_{15}$, placed above the slice, and a 15\,mm long rod of
Al$_{75}$Ni$_{25}$, placed below the slice. This configuration allows the
development of an error function shaped chemical interdiffusion profile
simultaneously to the development of a Gauss function shaped self--diffusion
profile. In a first approximation the diffusion of the enriched stable
isotope takes place at the mean concentration Al$_{80}$Ni$_{20}$
without influence of the changing chemical composition of the melt in
the diffusion zone. The only necessary correction results from the mass
spectrometric measurement of the self--diffusion profile. Here the measured
isotope incidences $i(^{62}{\rm Ni})$ of $^{62}$Ni have to be corrected
for the overlaying chemical concentration profile of natural Ni, $c_{\rm Ni}$, 
by using the following formula:
\begin{equation}
c(^{62}{\rm Ni}) = c_{\rm Ni} \left( i(^{62}{\rm Ni}) - i(^{62}{\rm Ni}^{0})
\right)
\end{equation}
with $i(^{62}{\rm Ni}^{0})$ the natural incidence of $^{62}$Ni and
$c(^{62}{\rm Ni})$ the concentration of this Ni isotope with respect to
all Ni isotopes. Typical concentration profiles of a diffusion experiment
are given in Fig.~\ref{fig1}.

The diffusion couple configuration minimizes the risk of
convection compared to conventional self--diffusion experiments
in pure melts because of the solutal stabilized density profile
of the melt column. This stabilizing effect has been described in
Refs.~\cite{garandet95,barrat96}. In a standard self--diffusion experiment
without chemical gradient the solutal stabilization effect is only due
to the enrichment of a tracer.

As a test for other mass transport processes we measured the mean--square
penetration depth $\bar{x}^2$ of interdiffusion as a function of time $t$.
We found a deviation from the linear behavior $\bar{x}^2=2 D_{\rm AB}t$.
This has been identified as sedimentation of Al$_{3}$Ni$_{2}$ during
solidification of the diffusion sample. This additional mass transport
was simply corrected by subtracting this contribution as an off--set
of the measured total mass transport.  This procedure adds a 5-10\%
error to the uncertainty of the diffusion coefficient. The total error
in the long--capillary measurements of the self-- and interdiffusion
coefficients is about 30--40\%.

\subsection{Neutron scattering experiments}
The second experimental technique used in this work is quasielastic
neutron scattering. In this case, the Al$_{80}$Ni$_{20}$ alloy was
prepared by arc melting of pure elements under a purified Argon
atmosphere.  The measurements were done at the time--of--flight
spectrometer IN6 of the Institut Laue-Langevin.  The standard Nb resistor
high temperature vacuum furnace of the ILL exhibits a temperature gradient
over the entire sample at 1800\,K that was less than five degrees and a
temperature stability within one degree.  For the scattering experiment
we used a thin--walled Al$_2$O$_3$ container that provides a hollow
cylindrical sample geometry of 22\,mm in diameter and a sample wall
thickness of 1.2\,mm.

\begin{figure}
\vspace*{0.2cm}
\includegraphics[width=\figurewidth]{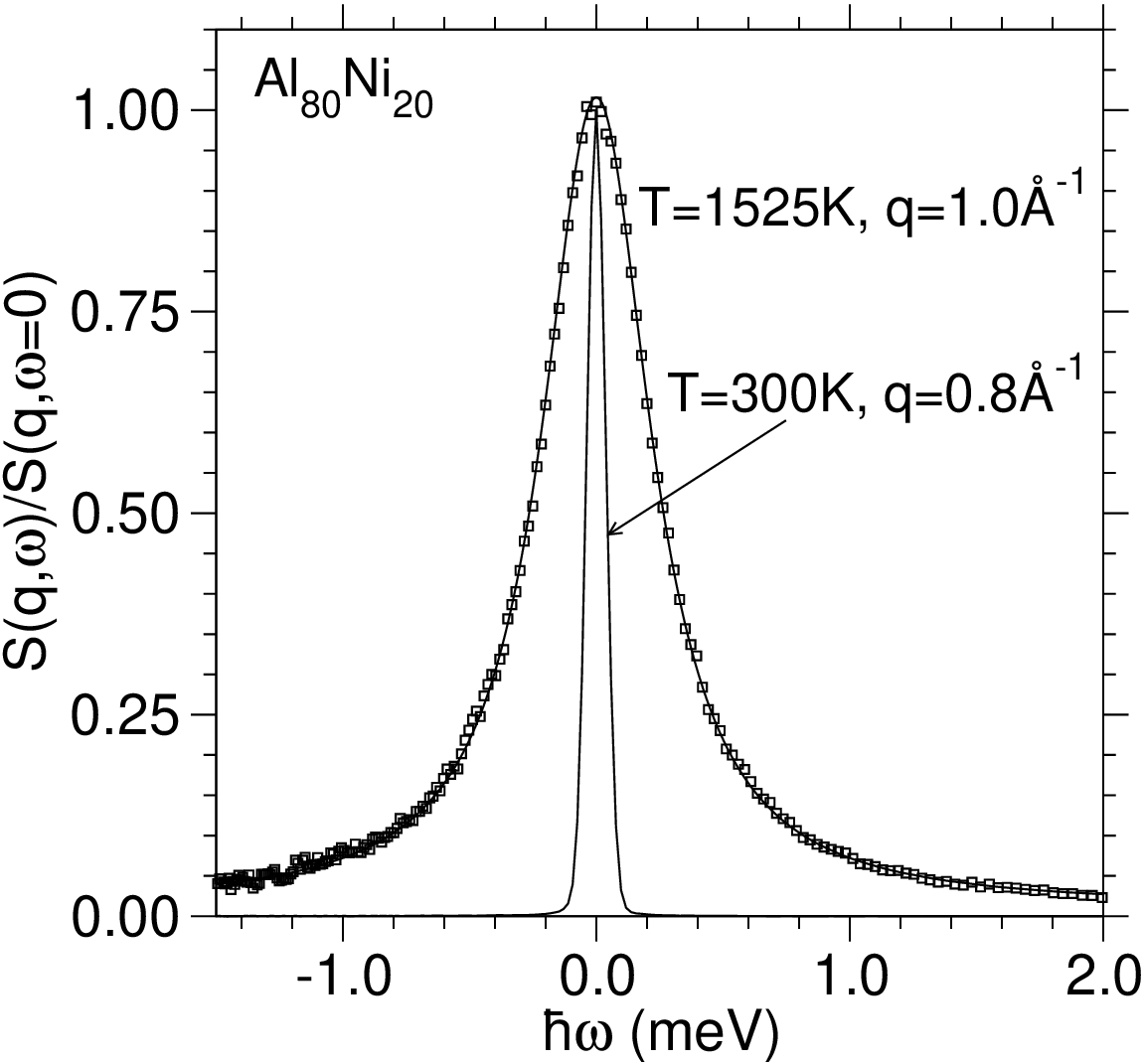}
\caption{\label{fig2}
Normalized scattering law of liquid Al$_{80}$Ni$_{20}$. The data at
300\,K represent the instrumental energy resolution function. The line is
a fit with a Lorentzian function that is convoluted with the instrumental
energy resolution function.  Diffusive motion of the atoms leads to a
broadening of the quasielastic signal from which the Ni self--diffusivity
can be obtained on an absolute scale.}
\end{figure}
An incident neutron wavelength of $\lambda\!=\!5.1\,\mbox{\AA}$
yielded an energy resolution of $\delta E\simeq92\,\mu\mbox{eV}$
(FWHM) and an accessible wave number range at zero energy transfer
of $q=0.4-2.0\,\mbox{\AA}^{-1}$.  Measurements were performed at
1350\,K, 1525\,K, 1670\,K and 1795\,K in 2 hour runs each. A run at
room temperature provided the instrumental energy resolution function.
The scattering law $S(q,\omega)$ was obtained by normalization to a
vanadium standard, accompanied by a correction for self absorption and
container scattering, and interpolation to constant wave numbers $q$.
Further, $S(q,\omega)$ was symmetrized with respect to the energy transfer
$\hbar\omega$ by means of the detailed balance factor.

Figure \ref{fig2} displays $S(q,\omega)$ at $q=1.0\,\mbox{\AA}^{-1}$
of liquid Al$_{80}$Ni$_{20}$ at 1525\,K and the crystalline alloy at
300\,K at $q=0.8\,\mbox{\AA}^{-1}$.  Diffusive motion in the liquid
leads to a broadening of the quasielastic signal.  The data were fitted
with an Lorentzian function that is convoluted with the instrumental
energy resolution function.  From the full width at half maximum of the
quasielastic line $\Gamma$ a $q$--dependent diffusion coefficient $D(q)$
can be computed via $D(q) = \Gamma / (2 \hbar q^2)$.  Towards small
$q$ incoherent scattering on the Ni atoms dominates the signal and
the diffusion coefficient $D(q)$ becomes constant yielding an estimate
of $D_{\rm s, Ni}$. Thus, the self--diffusion constant $D_{\rm s, Ni}$
can be determined on an absolute scale \cite{Mey02,MaMK04}.

\section{Details of the simulation}
\label{sec4}
For the computer simulations of the binary system Al$_{80}$Ni$_{20}$,
we used a potential of the embedded atom type that was recently derived
by Mishin {\it et al.}~\cite{mishin02}.  In a recent publication
\cite{das05}, we have shown that this potential reproduces very well
structural properties and the self--diffusion constant of Al--Ni melts at
various compositions. The present simulations are performed in a similar
way as the ones in the latter work: Systems of $N=1500$ particles 
($N_{\rm Ni}=300$, $N_{\rm Al}=1200$) are put in a cubic simulation
box with periodic boundary conditions. First, standard Monte--Carlo (MC)
simulations in the $NpT$ ensemble~\cite{binder_book} were used to fully
equilibrate the systems at zero pressure and to generate independent
configurations for MD simulations in the microcanonical ensemble.
In the latter case, Newton's equations of motion were integrated with the
velocity Verlet algorithm using a time step of 1.0\,fs at temperatures
$T\ge1500$\,K and 2.5\,fs at lower temperatures.

The masses were set to 26.981539\,amu and 58.69\,amu for aluminum and nickel,
respectively. At each temperature investigated, we made sure that the
duration of the equilibration runs exceeded the typical relaxation times of
the system.  The temperatures considered were 4490\,K, 2994\,K, 2260\,K,
1996\,K, 1750\,K, 1496\,K, 1300\,K, 1100\,K, 998\,K, 940\,K, 893\,K,
847\,K, 810\,K, 777\,K, 754\,K, 735\,K, 715\,K, 700\,K, 680\,K, and
665\,K.  In order to improve the statistics of the results we averaged at
each temperature over eight independent runs. At the lowest temperature,
the duration of the microcanonical production runs were 40 million time
steps, thus yielding a total simulation time of about 120\,ns. The
latter production runs were used to study the tagged particle dynamics.
For the calculation of the interdiffusion constant $D_{\rm AB}$ additional
production runs were performed in the temperature range $4490\,{\rm
K}\ge T \ge 715$\,K that extended the production runs for the tagged
particle dynamics by about a factor of ten. This was necessary in order
to obtain a reasonable statistics for $D_{\rm AB}$. Note that $D_{\rm
AB}$ is a collective quantity that does not exhibit the self--averaging
properties of the self--diffusion constant and thus it is quite demanding
to determine transport coefficients such as the interdiffusion constant
or the shear viscosity from a MD simulation.

\section{Results}
\label{sec5}
In Eq.~(\ref{darken}), the interdiffusion constant $D_{\rm AB}$ is
expressed as a linear combination of the self--diffusion constants. The
prefactor in this formula is a product of the thermodynamic factor $\Phi$
and the Manning factor $S$. Whereas $\Phi$ can be computed from structural
input, the Manning factor contains the collective dynamic correlations in
the expression for $D_{\rm AB}$ (see Sec.~\ref{sec2}). In the following,
we compare the simulated diffusion constants for Al$_{80}$Ni$_{20}$ to
those from experiments. Moreover, the simulations are used to disentangle
differences between self--diffusion constants and the interdiffusion
constants with respect to the thermodynamic quantity $\Phi$ and the
dynamic quantity $S$.

\begin{figure}
\vspace*{0.5cm}
\includegraphics[width=0.8\figurewidth]{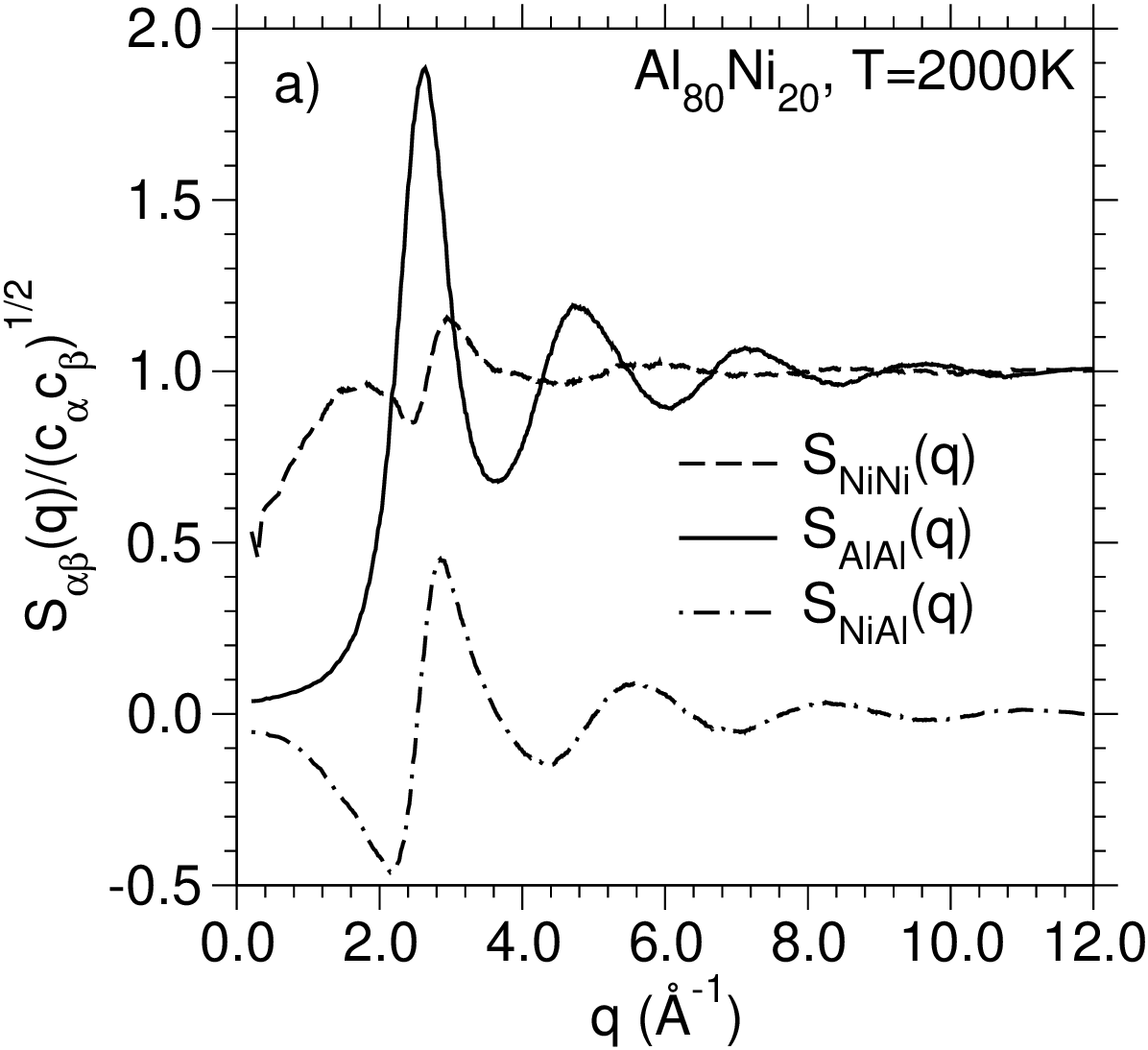}
\includegraphics[width=0.8\figurewidth]{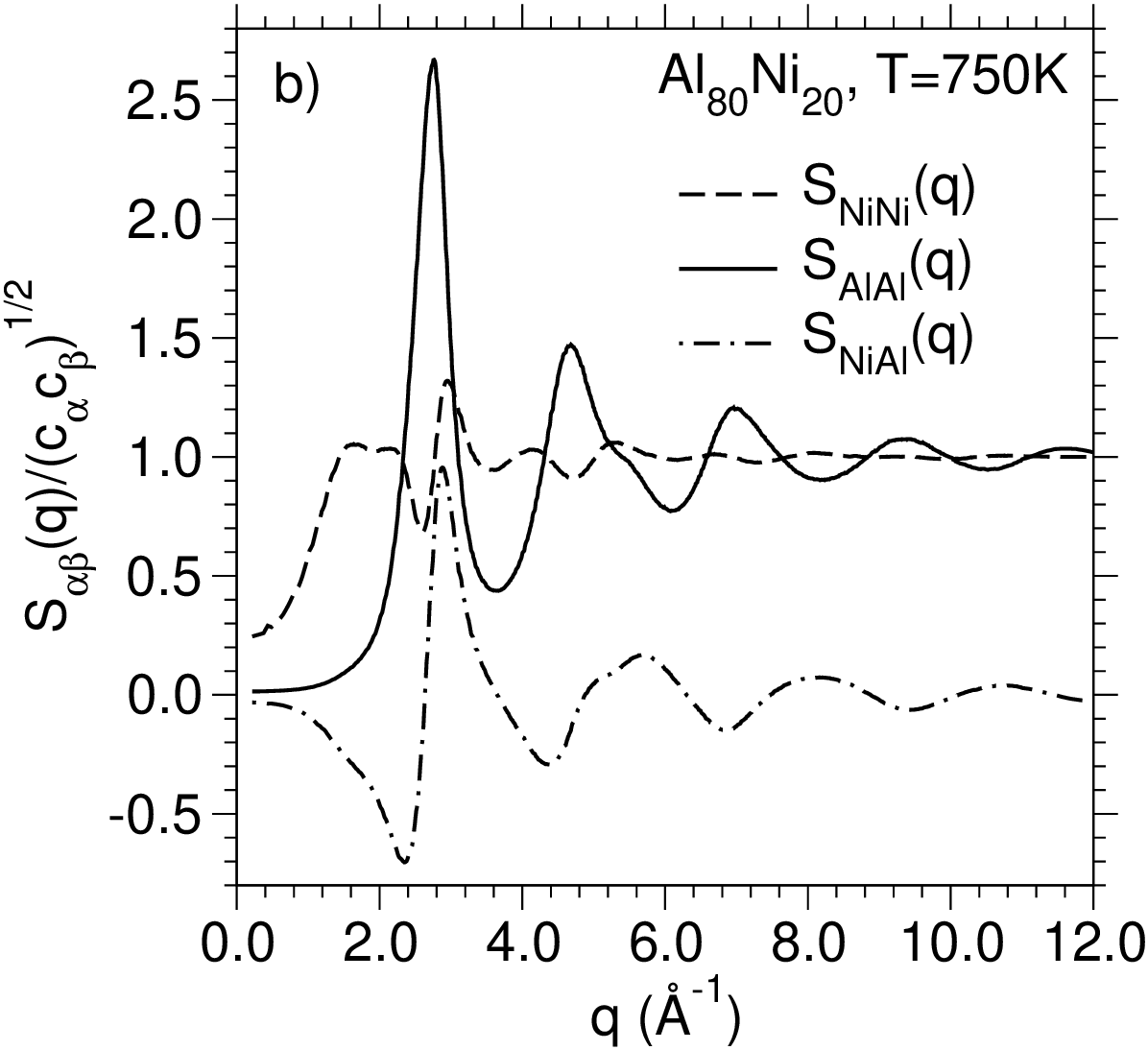}
\caption{\label{fig3}
Partial structure factors, as obtained from the 
MD simulation, for a) $T=2000$\,K and b) $T=750$\,K. The multiplication
by $1/(c_{\alpha} c_{\beta})^{1/2}$ is introduced to increase the amplitude
of $S_{\rm NiNi}(q)$ relative to that of $S_{\rm AlAl}(q)$. Note that
the factor $1/(c_{\alpha} c_{\beta})^{1/2}$ leads also to the asymptotic
value $S_{\alpha\alpha}(q)=1$ for $q\to\infty$.}
\end{figure}
First, we discuss static structure factors at different temperatures,
as obtained from the MD simulation. Figure \ref{fig3} displays the
different partial structure factors at the temperatures $T=2000$\,K and
$T=750$\,K. At both temperatures, a broad prepeak around the wavenumber
$q=1.8$\,\AA$^{-1}$ emerges in the NiNi correlations, which indicates
the presence of chemical short--ranged order (CSRO). This feature is
absent in the AlAl correlations.  In a recent work \cite{das05}, we
have found that the prepeak in $S_{\rm NiNi}(q)$ is present in a broad
variety of Al--Ni compositions, ranging from $x_{\rm Ni}=0.1$ to $x_{\rm
Ni}=0.9$. However, the width of the prepeak decreases significantly with
increasing Ni concentration and, in melts with a high Ni concentration,
it appears also in $S_{\rm AlAl}(q)$. The prepeak in $S_{\alpha \beta}(q)$
describes repeating structural units involving next--nearest $\alpha
\beta$ neighbors which are built in inhomogeneously into the structure.
Of course, for the Al rich system Al$_{80}$Ni$_{20}$ considered in this
work, only next--nearest Ni--Ni units exhibit the CSRO that is reflected
in the prepeak.

\begin{figure}
\vspace*{0.5cm}
\includegraphics[width=0.8\figurewidth]{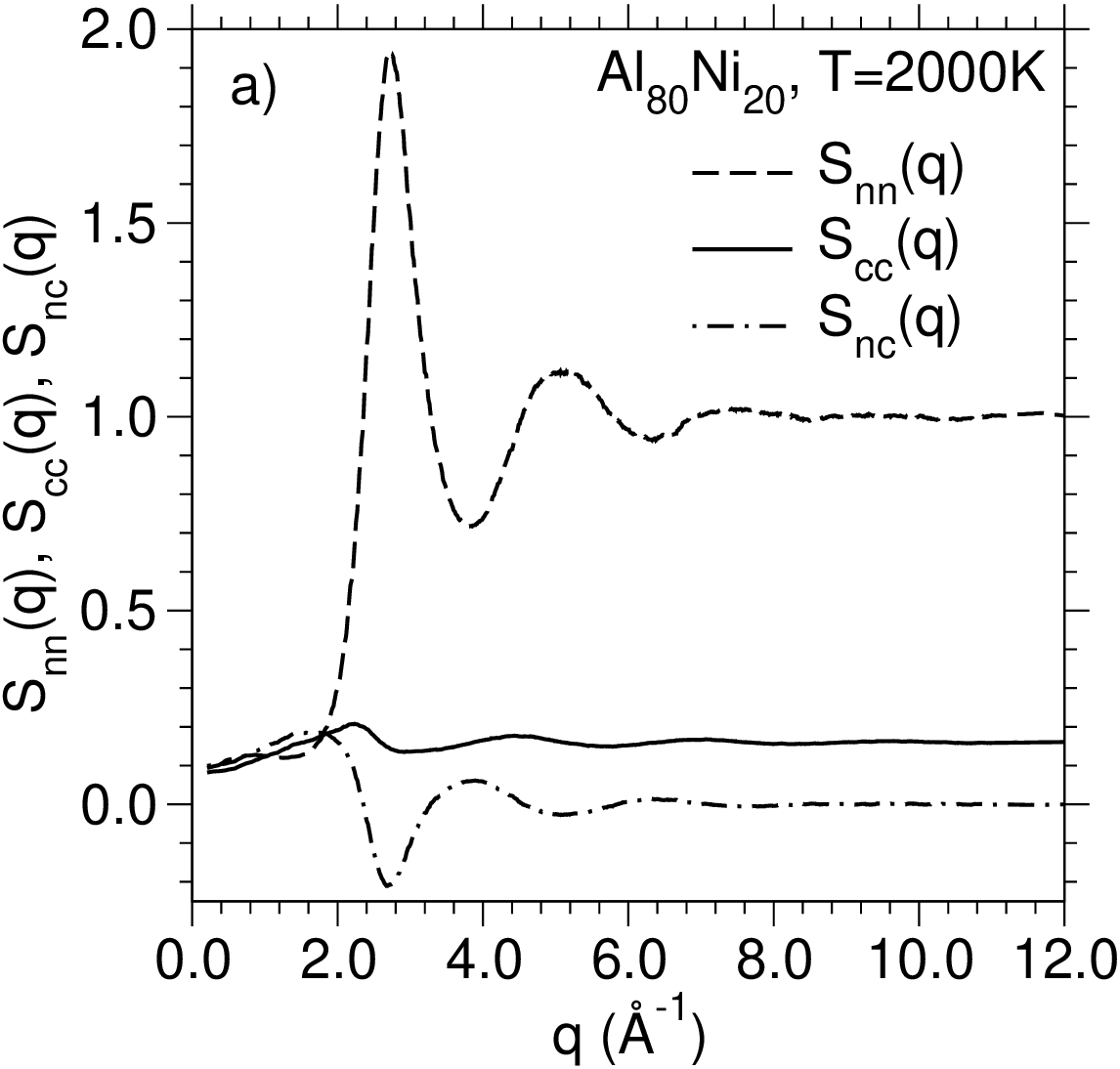}
\includegraphics[width=0.8\figurewidth]{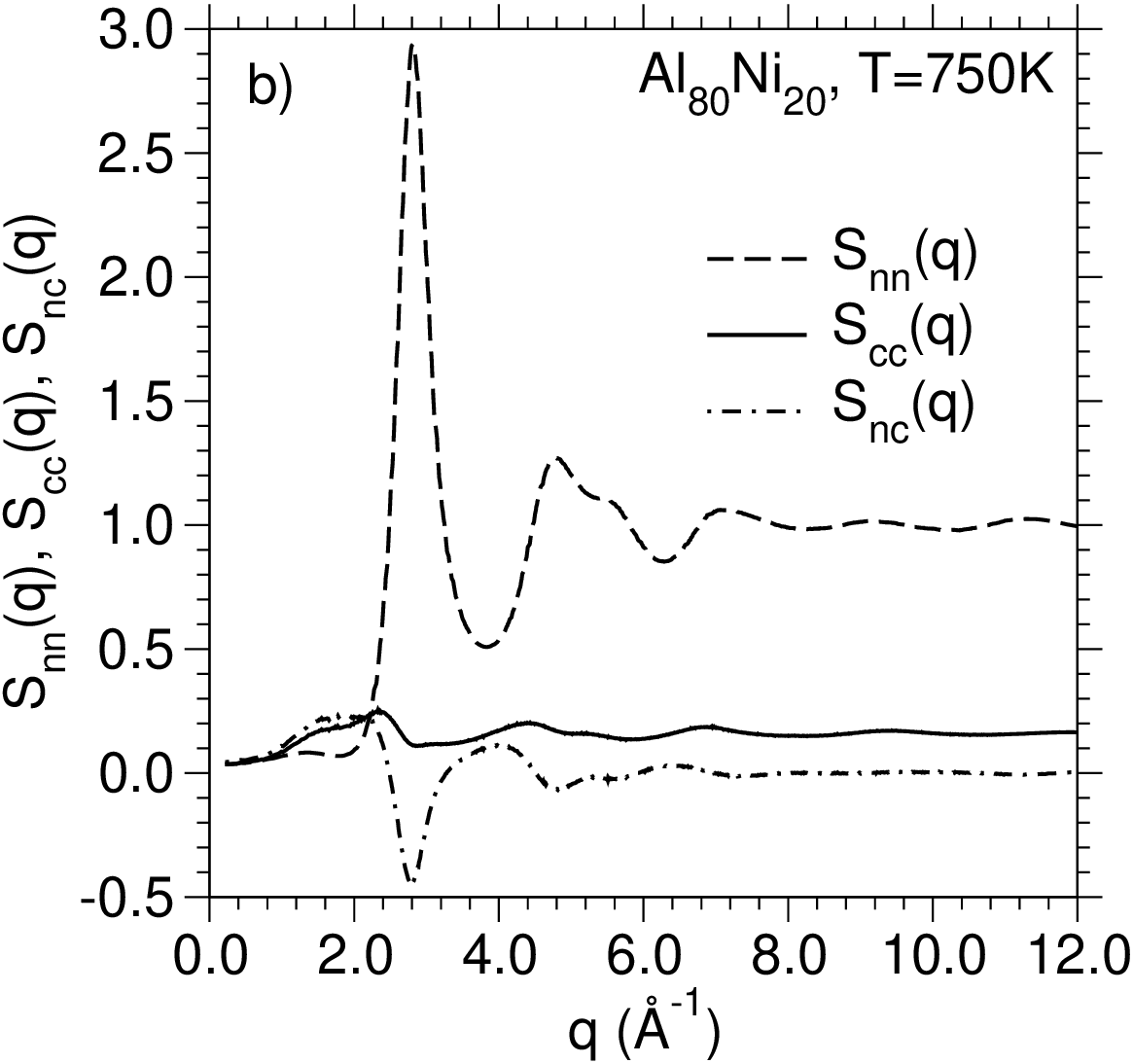}
\caption{\label{fig4}
Bhatia--Thornton structure factors, as obtained from the MD simulation, 
at a) $T=2000$\,K and b) $T=750$\,K.}
\end{figure}
From the partial static structure factors, the Bhatia--Thornton structure
factors can be determined according to Eqs.~(\ref{eq_scc}), (\ref{eq_snn})
and (\ref{eq_snc}). These quantities are shown in Fig.~\ref{fig4}, again
at $T=2000$\,K and at $T=750$\,K. Although these structure factors look
very different for $q>2$\,\AA$^{-1}$, they are essentially identical
in the limit $q\to 0$.  As we have indicated before, the static
susceptibility, associated with concentration fluctuations, can be
extracted from the structure factor $S_{cc}(q)$ in the limit $q\to0$.
As we can infer from Fig.~\ref{fig4}, at the temperature $T=750$\,K
the value of this susceptibility is very small. The small value of
$S_{cc}(q=0)$ reveals that concentration fluctuations on large length
scales are strongly suppressed. This is the typical behavior of a dense
fluid that exhibits a strong ordering tendency.  In contrast, at a
critical point of a demixing transition a divergence of $S_{cc}(q=0)$
is expected.

\begin{figure}
\vspace*{0.3cm}
\includegraphics[width=\figurewidth]{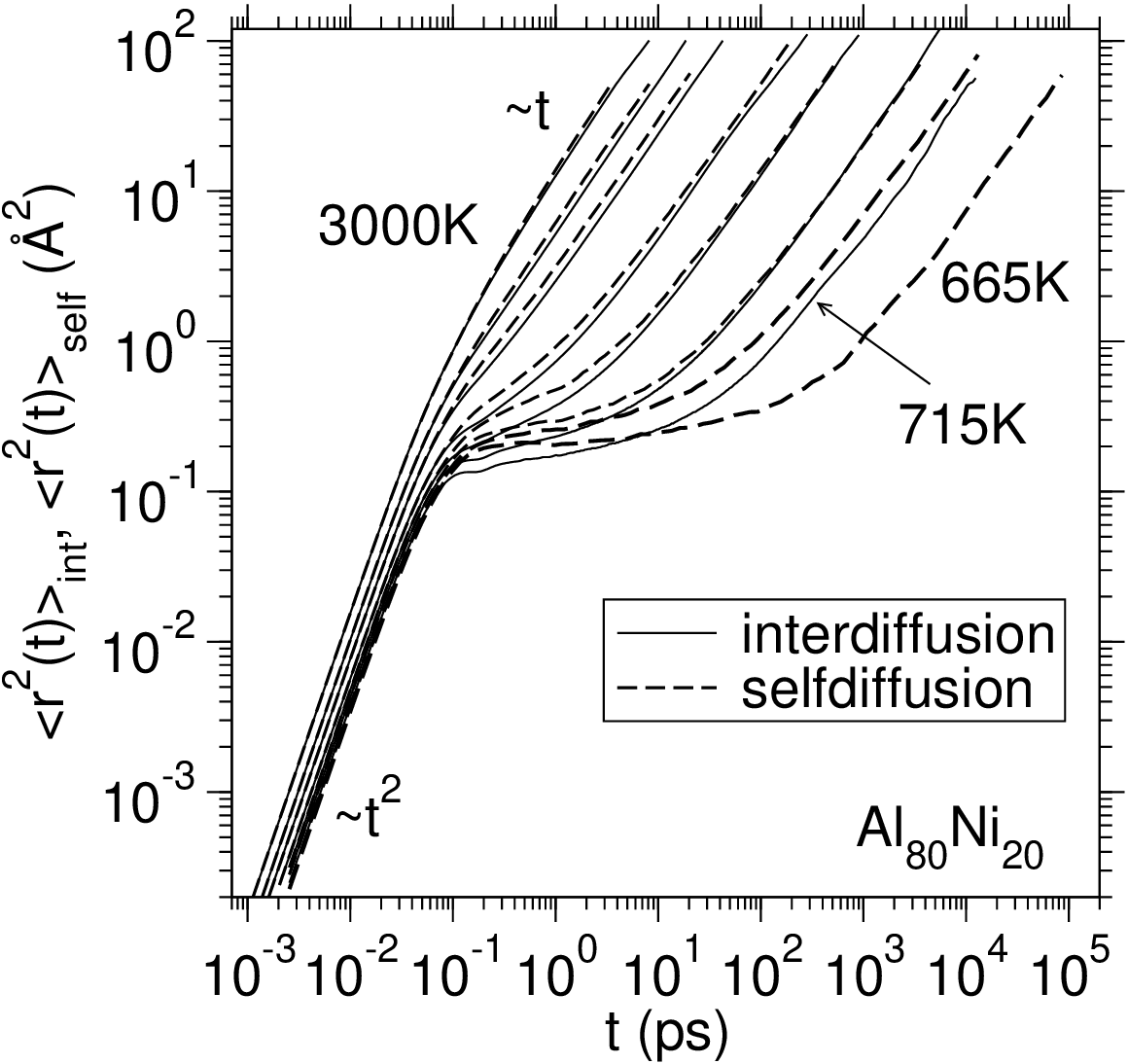}
\caption{\label{fig5}
Simulation results of mean squared displacements (MSD) for self--diffusion
(dashed lines) and interdiffusion (solid lines) for the
temperatures $T=3000$\,K, 2000\,K, 1500\,K, 1000\,K,
850\,K, 750\,K, 715\,K, and 665\,K (corresponding to the curves
from left to right. Note that for $T=665$\,K only $\langle r^2(t) \rangle_{\rm self}$
was calculated. For the definitions of the MSD's see Eqs.~(\ref{eq_msdint})
and (\ref{eq_msdself}).}
\end{figure}
As we have seen in Sec.~\ref{sec2}, the ratio $D_{\rm AB}/\Phi$
can be expressed as a linear combination of the self--diffusion 
constants, provided $S=1$ holds. In order to quantify the
temperature dependence of $S$, we first define the following 
mean--squared displacements:
\begin{eqnarray}
  \langle r^2(t) \rangle_{\rm int} & = &
    \left( 1+ \frac{m_{\rm A}c_{\rm A}}{m_{\rm B}c_{\rm B}} \right)^2
   N c_{\rm A} c_{\rm B} \times \nonumber \\
 & & \times \langle 
    \left[{\bf R}_{\rm A}(t) - {\bf R}_{\rm A}(0)\right]^2 \rangle 
      \label{eq_msdint} \\
   \langle r^2(t) \rangle_{\rm self} & = & 
       c_{\rm A} \frac{1}{N_{\rm B}} \sum_{j=1}^{N_{\rm B}} 
      \langle \left[ {\bf r}_j^{\rm (B)} (t) - {\bf r}_j^{\rm (B)} (0) \right]^2 \rangle + 
        \nonumber \\
    & & c_{\rm B} \frac{1}{N_{\rm A}} \sum_{j=1}^{N_{\rm A}} 
      \langle \left[ {\bf r}_j^{\rm (A)} (t) - {\bf r}_j^{\rm (A)} (0) \right]^2 \rangle
      \label{eq_msdself}
\end{eqnarray}
Whereas the interdiffusion constant can be calculated via $D_{\rm AB}=
{\rm lim}_{t\to\infty} \Phi \langle r^2(t) \rangle_{\rm int}/(6t)$,
the equation $D_{\rm AB}= {\rm lim}_{t\to\infty} \Phi \langle r^2(t)
\rangle_{\rm self}/(6t)$ is only correct for $S=1$. Figure \ref{fig5}
shows the quantities $\langle r^2(t) \rangle_{\rm int}$ and $\langle
r^2(t) \rangle_{\rm self}$ for the different temperatures.  Both MSD's
show a very similar behavior. At high temperature, a crossover from
a ballistic regime ($\propto t^2$) at short times to a diffusive
regime ($\propto t$) at long times can be seen. At low temperature, a
plateau--like region develops at intermediate times, i.e.~between the
ballistic and the diffusive regime.  With decreasing temperature, the
plateau becomes more pronounced. In $\langle r^2(t) \rangle_{\rm self}$,
the plateau indicates the so--called cage effect \cite{glassbook}.
The tagged particle is trapped by its neighbors on a time scale
that increases with decreasing temperature. Although the MSD for
the interdiffusion, $\langle r^2(t) \rangle_{\rm int}$, describes
also collective particle transport, the plateau in this quantity has
the same origin: The particles are ``arrested'' on intermediate time
scales. Moreover, the differences between $\langle r^2(t) \rangle_{\rm
self}$ and $\langle r^2(t) \rangle_{\rm int}$ are anyway very small in
the whole time and temperature range under consideration. This means that
the cross correlations do not give a large contribution to $\langle r^2(t)
\rangle_{\rm int}$.

\begin{figure}
\includegraphics[width=0.8\figurewidth]{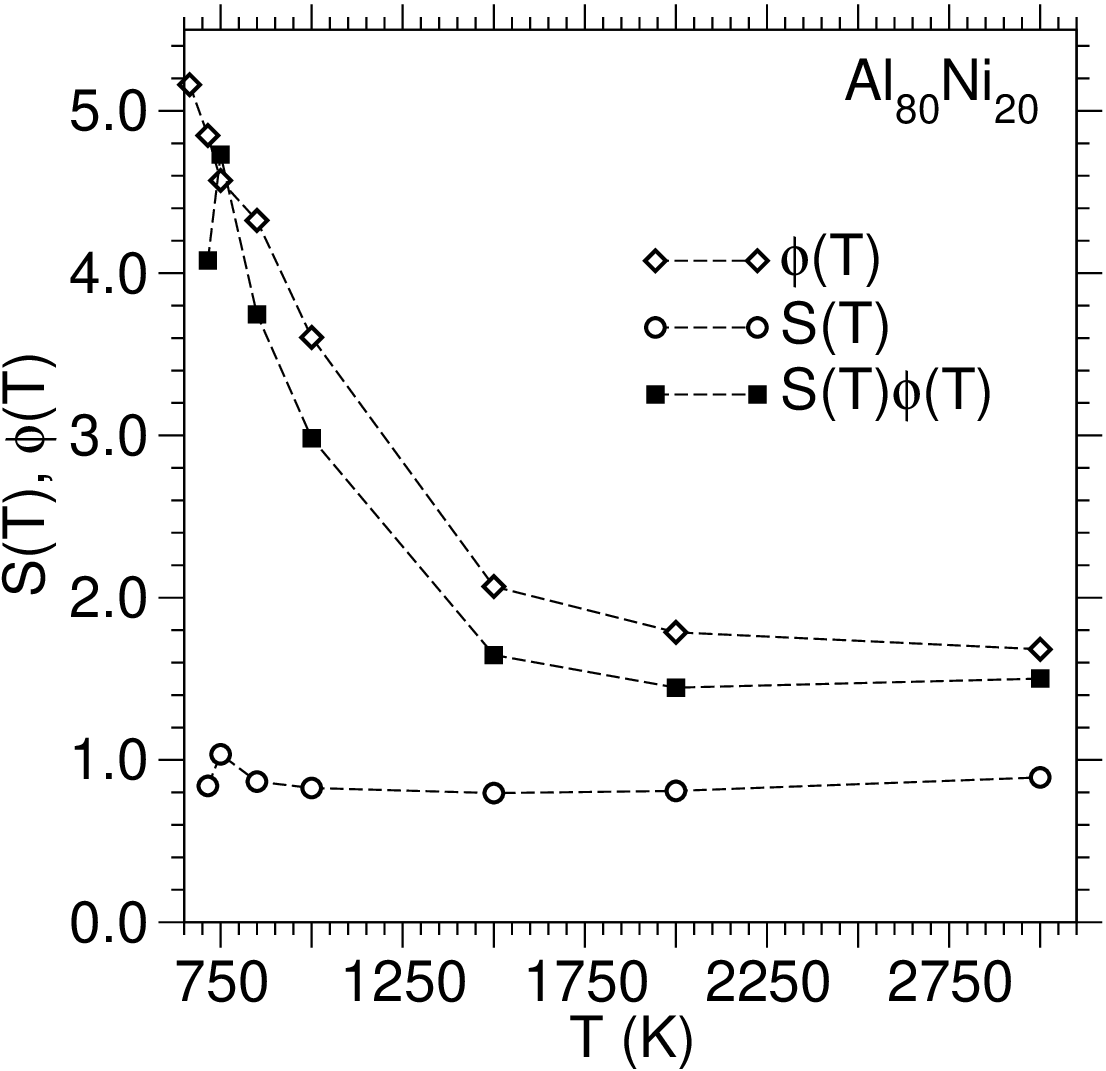}
\caption{\label{fig6}
Thermodynamic factor $\phi$, ``Manning'' factor $S(T)$, and the product
of both as obtained from the simulation.}
\end{figure}
From the MSD's in Fig.~\ref{fig5}, the Manning factor $S$ can be
extracted using Eq.~(\ref{eq_manning}). In Fig.~\ref{fig6} we see that
the Manning factor varies only slightly over the whole temperature range,
located around values between 0.8 and 1.0.
Also shown in Fig.~\ref{fig6} is the
thermodynamic factor $\Phi$ and the product $\Phi S$. We have extracted
$\Phi$ from the extrapolation of the structure factors $S_{cc}(q)$
toward $q\to 0$ [see Eq.~(\ref{eq_phi})].  In contrast to the Manning
factor $S$, the thermodynamic factor $\Phi$ increases significantly
with decreasing temperature and thus, also the change in the product
$\Phi S$ is dominated by the change in $\Phi$.  Therefore, differences
in the qualitative behavior between the self--diffusion constants and the
interdiffusion constant are dominated by the thermodynamic factor.

\begin{figure}[t]
\vspace{0.2cm}
\includegraphics[width=0.92\figurewidth]{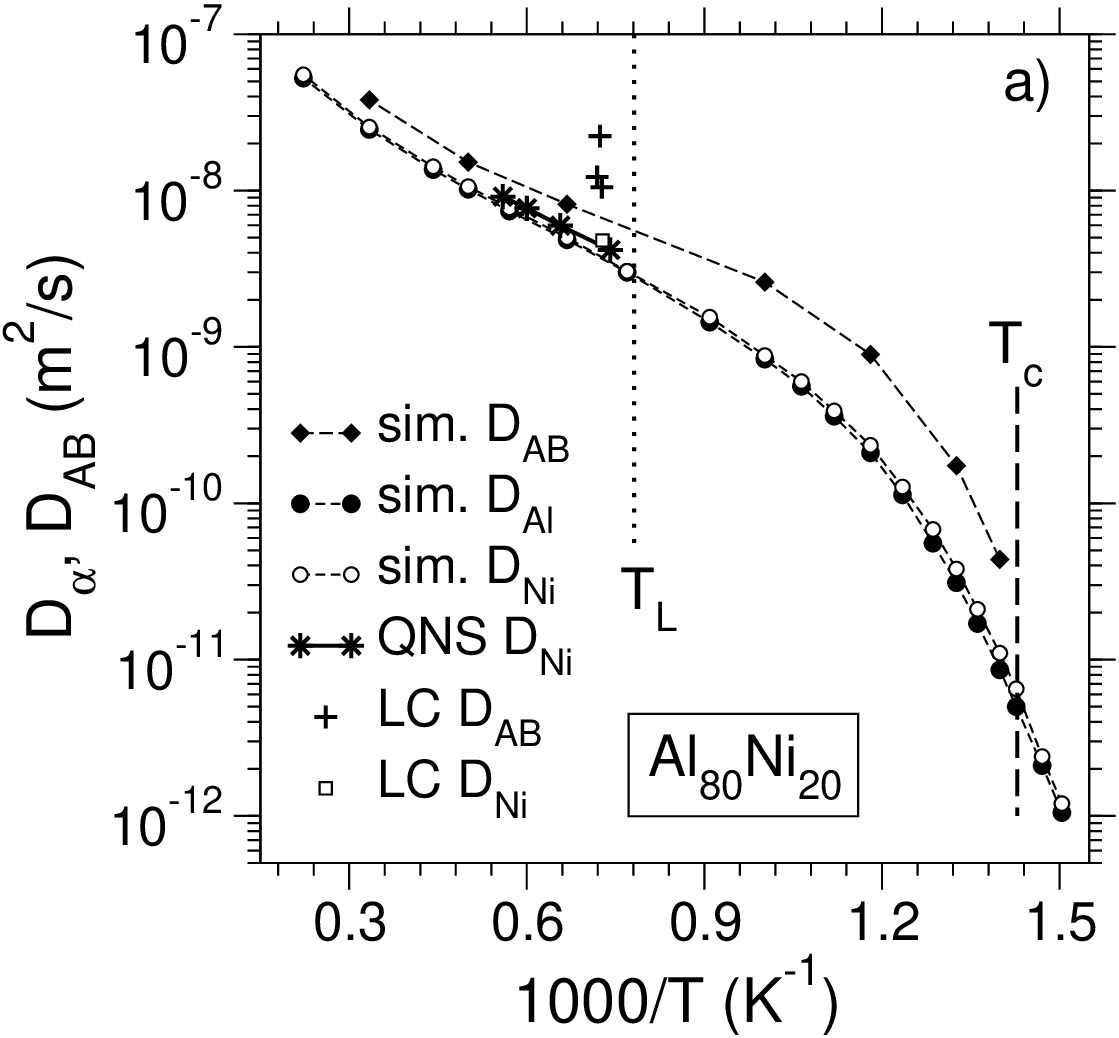}
\includegraphics[width=0.92\figurewidth]{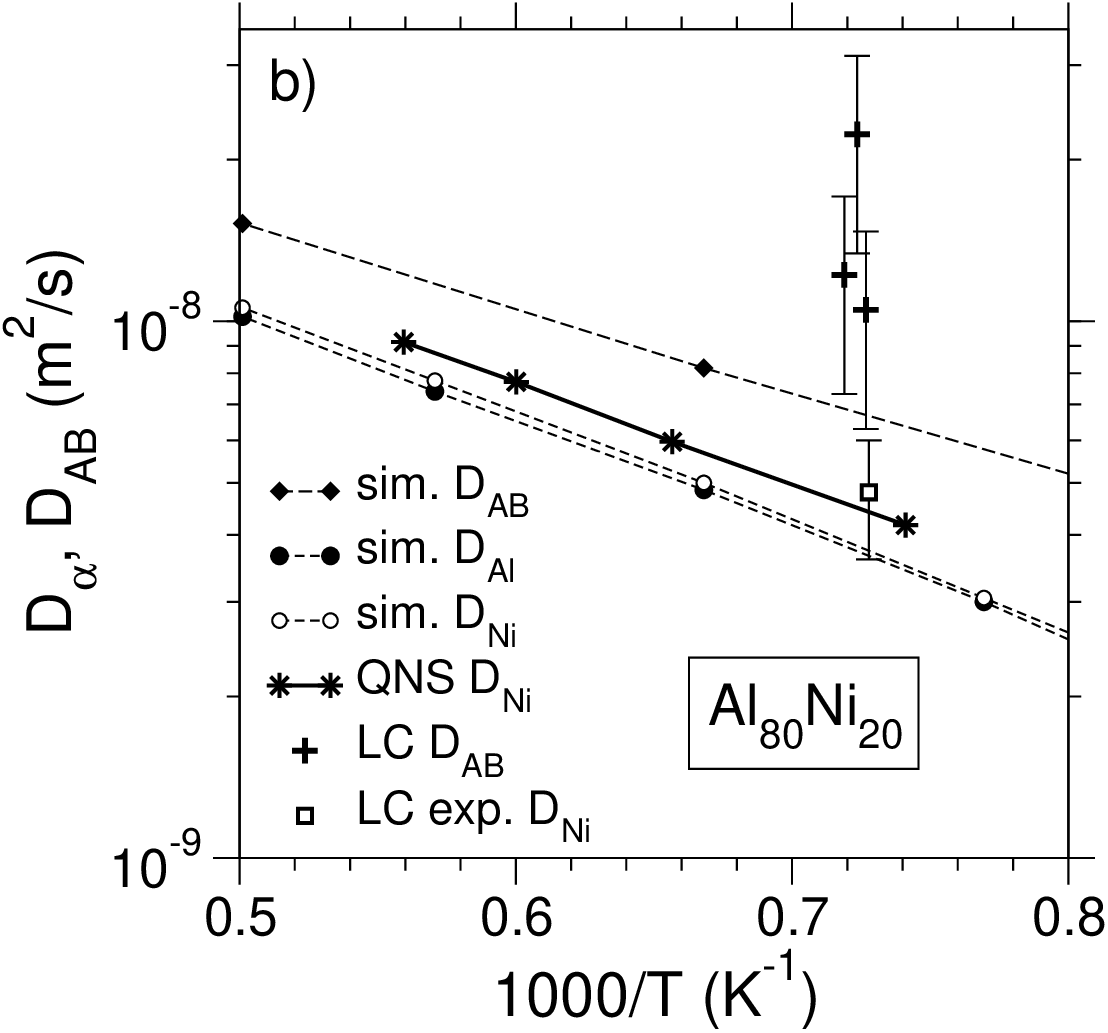}
\caption{\label{fig7}
Arrhenius plot of interdiffusion and self--diffusion constants, as obtained
from experiment and simulation, as indicated. The experimental results are
measured by quasielastic neutron scattering (QNS) and by the LC technique.
The lines through the data points are guides to the eye. The vertical
dotted line in a) marks the location of the experimental liquidus
temperature, $T_{\rm L}\approx1280$\,K. The vertical dashed line is at
the location of the critical temperature of mode coupling theory, $T_{\rm
c}\approx 700$\,K, as estimated by the MD simulation \cite{das07}.  Panel
b) is an enlargement of the data of panel a) in a temperature range above
$T_{\rm L}$. The error bars of simulation and QNS data are of the order 
of the size of the symbols.}
\end{figure}
An Arrhenius plot of the diffusion constants as obtained from simulation
and experiment is shown in Fig.~\ref{fig7}.  The self--diffusion constants
$D_{\rm Ni}$ and $D_{\rm Al}$ from the simulation are very similar over
the whole temperature range 4490\,K$\ge T \ge 665$\,K. In a recent
publication \cite{das05}, we have found that, in the framework of
our simulation model, this similarity of the self--diffusion constants
occurs in Al rich compositions of the system Al--Ni, say for $c_{\rm
Al}>0.7$. Whether this is also true in real systems is an open
question. However, the neutron scattering results for $D_{\rm Ni}$
as well as the single point obtained from the LC measurement
is in very good agreement with the simulation data. 

Asta {\it et al.}~\cite{asta99} have computed the concentration dependence
of the self--diffusion constants at $T=1900$~K using two different
embedded atom potentials, namely the one proposed by Voter and Chen
\cite{voter78} and the one proposed by Foiles and Daw \cite{foiles87}.
For both potentials, they find very similar values for $D_{\rm Ni}$
and $D_{\rm Al}$ in Al$_{80}$Ni$_{20}$, in agreement with our results.
However, their results for the Ni diffusion constant are significantly
higher than the ones found in our quasielastic neutron scattering experiment
and our simulation. They report the values $D_{\rm Ni}\approx 1.5 \cdot
10^{-8}$\,m$^2$/s and $D_{\rm Ni}\approx 1.9 \cdot 10^{-8}$\,m$^2$/s for
the Voter--Chen potential and the Foiles--Daw potential, respectively,
whereas we obtain $D_{\rm Ni}\approx 10^{-8}$\,m$^2$/s from simulation
and experiment. Thus, the potential proposed by Mishin {\it et al.}
\cite{mishin02}, which is used in this work, leads to a better agreement
with the experiment, as far as self--diffusion in Al$_{80}$Ni$_{20}$
is concerned.

We emphasize that the statistical error in both the neutron scattering
data and the simulation data for the self--diffusion constants is relatively
small. In both cases, the error bars for the corresponding data points
in Fig.~\ref{fig7} are smaller than the size of the symbols.

Due to the lack of self--averaging, it is much more difficult to yield
accurate results for $D_{\rm AB}$ from the simulation. Therefore, in this
case we considered a smaller temperature range than for the self--diffusion
constants to yield results with reasonable accuracy. As we can infer
from Fig.~\ref{fig7}, the interdiffusion constant is larger than the
self--diffusion constants over the whole temperature range. The difference
becomes more pronounced with decreasing temperature. At $T=715$\,K,
the diffusion coefficient $D_{\rm AB}$ is about a factor of 3 larger
than $D_{\rm Ni}$ and $D_{\rm Al}$.  This behavior is of course due
to the increase of the thermodynamic factor $\Phi$ at low temperature.
Also included in Fig.~\ref{fig7} are the results of the LC measurements
of $D_{\rm AB}$ and $D_{\rm Ni}$. These results are much less accurate
than those of the quasielastic neutron scattering experiments for the
determination of $D_{\rm Ni}$ (see the error bars for the LC data
in Fig.~\ref{fig7}b).  Nevertheless, the LC data show that $D_{\rm
AB}>D_{\rm Ni}$ holds, in agreement with the simulation results.

\section{Conclusion}
\label{sec6}

A combination of experiment and molecular dynamics (MD) simulation
has been used to investigate the diffusion dynamics in liquid
Al$_{80}$Ni$_{20}$. We find good agreement between simulation and
experiment. Both in experiment and in simulation, the interdiffusion
constant is higher than the self--diffusion constants.  This is valid in
the whole temperature range considered in this work, i.e.~in the normal
liquid state as well as in the undercooled regime.  In the latter regime
(which is only accessible by the simulation), the difference between
the interdiffusion constant and the self--diffusion constants increases
with decreasing temperature.

All these observations can be clarified by the detailed information
provided by the MD simulation.  Both the thermodynamic factor
$\Phi$ and the Manning factor $S$ have been estimated directly and
accurately over a wide temperature range, as well as self--diffusion and
interdiffusion coefficients.  The central result of this work is shown
in Fig.~\ref{fig6}.  Whereas the thermodynamic factor $\Phi$ increases
significantly by lowering the temperature, the Manning factor $S$ shows
only a weak temperature dependence. Moreover, the value of $S$ is close
to one which means that dynamic cross correlations are almost negligible
and thus, even in the undercooled regime, the Darken equation is a good
approximation. The temperature dependence of $\Phi$ is plausible for a
dense binary mixture with a strong ordering tendency. The situation is
similar to the case of the isothermal compressibility which normally
decreases with temperature in a densely packed liquid leading to very
low values in the undercooled regime. In the same sense, the response to
a macroscopic concentration fluctuation described by $S_{\rm cc}(q=0)$
tends to become smaller and smaller towards the undercooled regime which
corresponds to an increase of $\Phi$ with decreasing temperature (since
$\Phi\propto 1/S_{cc}(q=0)$).

We note that the data shown for $D_{\rm AB}$ are all above the critical
temperature $T_{\rm c}$ of mode coupling theory which is around 700\,K
for our simulation model (see Fig.~\ref{fig7}) \cite{das07}. Since
it is expected that the transport mechanism changes below $T_{\rm c}$
\cite{glassbook}, it would be interesting to see how such a change in the
transport mechanism is reflected in the interdiffusion constants. This
issue is the subject of forthcoming studies.

\begin{acknowledgments}
We are grateful to Kurt Binder for stimulating discussions and a critical
reading of the manuscript.  We gratefully acknowledge financial support
within the SPP 1120 of the Deutsche Forschungsgemeinschaft (DFG)
under grants Bi314/18, Ma1832/3-2 and Me1958/2-3 and from DFG grant
Gr2714/2-1. One of the authors was supported through the Emmy Noether
program of the DFG, grants Ho2231/2-1/2 (J.H.).  Computing time on the
JUMP at the NIC J\"ulich is gratefully acknowledged.
\end{acknowledgments}


\references

%
\bibitem{hansen}
J.--P. Hansen and I.R. McDonald,
{\it Theory of Simple Liquids} (Academic Press, London, 1986).
%
\bibitem{shimoji}
M. Shimoji and T. Itami, 
{\it Atomic Transport in Liquid Metals}
(Trans. Tech. Publications, Aedermannsdorf, 1986). 
%
\bibitem{allnatt87}
A.R. Allnatt and A.B. Lidiard,
Rep. Prog. Phys. {\bf 50}, 373 (1987).
%
\bibitem{cahn83}
R.W. Cahn and P. Haasen (eds.),
{\it Physical metallurgy, Part 1 and 2}
(North-Holland, Amsterdam, 1983).
%
\bibitem{hohenberg77}
P.C. Hohenberg and B.I. Halperin,
Rev. Mod. Phys. {\bf 49}, 435 (1977).
%
\bibitem{glassbook}
K. Binder and W. Kob,
{\it Glassy Materials and Disordered Solids --- An Introduction to 
Their Statistical Mechanics}
(World Scientific, London, 2005).
%
\bibitem{vogelsang88}
R. Vogelsang and C. Hoheisel,
Phys. Rev. A {\bf 38}, 6296 (1988);
H.P. van den Berg and C. Hoheisel,
Phys. Rev. A {\bf 42}, 2090 (1990).
%
\bibitem{kehr89}
K.W. Kehr, K. Binder, and S.M. Reulein,
Phys. Rev. B {\bf 39}, 4891 (1989).
%
\bibitem{hess90}
W. Hess, G. N\"agele, and A.Z. Akcasu,
J. Polymer Sc. B {\bf 28}, 2233 (1990).
%
\bibitem{trullas94}
J. Trull\`as and J.A. Padr\`o,
Phys. Rev. E {\bf 50}, 1162 (1994).
%
\bibitem{baumketner99}
A. Baumketner and Ya. Chushak,
J. Phys.: Condens. Matter {\bf 11}, 1397 (1999).
%
\bibitem{wax07}
J.--F. Wax and N. Jakse,
Phys. Rev. B {\bf 75}, 024204 (2007).
%
\bibitem{darken49} 
L.S. Darken, 
Trans. AIME {\bf 180}, 430 (1949).
%
\bibitem{fuchs93}
M. Fuchs and A. Latz,
Physica A {\bf 201}, 1 (1993).
%
\bibitem{asta99}
M. Asta, D. Morgan, J.J. Hoyt, B. Sadigh, J.D. Althoff, D. de Fontaine,
and S.M. Foiles,
Phys. Rev. B {\bf 59}, 14271 (1999).
%
\bibitem{faupel03}
F. Faupel, W. Frank, M.--P. Macht, H. Mehrer, V. Naundorf, K. R\"atzke,
H.R. Schober, S.K. Sharma, and H. Teichler,
Rev. Mod. Phys. {\bf 75}, 237 (2003).
%
\bibitem{mishin02} 
Y. Mishin, M. J. Mehl, and D. A. Papaconstantopoulos,
Phys. Rev. B {\bf 65}, 224114 (2002).
%
\bibitem{das05}
S. K. Das, J. Horbach, M. M. Koza, S. Mavila Chatoth, and A. Meyer,
Appl. Phys. Lett. {\bf 86}, 011918 (2005). 
%
\bibitem{batalin83}
G.L. Batalin, E.A. Beloborodova, and V.G. Kazimirov,
{\it Thermodynamics and the Constitution of Liquid Al Based
Alloys} (Metallurgy, Moscow, 1983).
%
\bibitem{ayushina69}
G.D. Ayushina, E.S. Levin, and P.V. Geld,
Russ. J. Phys. Chem. {\bf 43}, 2756 (1969).
%
\bibitem{maret90}
M. Maret, T. Pomme, A. Pasturel, and P. Chieux,
Phys. Rev. B {\bf 42}, 1598 (1990).
%
\bibitem{saadeddine94}
S. Sadeddine, J.F. Wax, B. Grosdidier, J.G. Gasser, C. Regnaut,
and J.M. Dubois, 
Phys. Chem. Liq. {\bf 28}, 221 (1994).
%
\bibitem{asta01}
M. Asta, V. Ozolins, J.J. Hoyt, and M. van Schilfgaarde,
Phys. Rev. B {\bf 64}, 020201(R) (2001).
%
\bibitem{enthalpy}
V.S. Sudovtseva, A.V. Shuvalov, N.O. Sharchina,
Rasplavy No. {\bf 4}, 97 (1990);
U.K. Stolz, I. Arpshoven, F. Sommer, and B. Predel,
J. Phase Equilib. {\bf 14}, 473 (1993);
K.V. Grigorovitch and A.S. Krylov,
Thermochim. Acta {\bf 314}, 255 (1998).
%
\bibitem{das06}
S.K. Das, J. Horbach, and K. Binder,
J. Chem. Phys. {\bf 119}, 1547 (2003);
S.K. Das, J. Horbach, K. Binder, M.E. Fisher, J.V. Sengers,
J. Chem. Phys. {\bf 125}, 024506 (2006);
S.K. Das, M. E. Fisher, J.V. Sengers, J. Horbach, K. Binder,
Phys. Rev. Lett. {\bf 97}, 025702 (2006).
%
\bibitem{raineri89}
F.O. Raineri and H.L. Friedman,
J. Chem. Phys. {\bf 91}, 5633 (1989);
F.O. Raineri and H.L. Friedman,
J. Chem. Phys. {\bf 91}, 5642 (1989).
%
\bibitem{bhatia70} 
A.B. Bhatia and D.E. Thornton, 
Phys. Rev. B {\bf 52}, 3004 (1970).
%
\bibitem{holland06}
D. Holland--Moritz, O. Heinen, R. Bellissent, 
T. Schenk, and D.M. Herlach,
Int. J. Mat. Res. {\bf 97}, 948 (2006).
%
\bibitem{manning61} 
J.R. Manning, 
Phys Rev. {\bf 124}, 470 (1961).
%
\bibitem{allen94}
M. P. Allen, D. Brown, and A. J. Masters,
Phys. Rev. E {\bf 49}, 2488 (1994);
M. P. Allen,
Phys. Rev. E {\bf 50}, 3277 (1994).
%
\bibitem{griesche06}
A. Griesche, F. Garcia Moreno, M.P. Macht, and G. Froh\-berg,
Mat. Sc. Forum {\bf 508}, 567 (2006).
%
\bibitem{griesche04}
A. Griesche, M.P. Macht, J.P. Garandet, and G. Froh\-berg,
J Non--Cryst. Solids {\bf 336}, 173 (2004).
%
\bibitem{griesche07}
A. Griesche, M.P. Macht, and G. Frohberg,
unpublished.
%
\bibitem{garandet95}
J.P. Garandet, C. Barrat, and T. Duffar,
Int. J. Heat Mass Transfer {\bf 38}, 2169 (1995).
%
\bibitem{barrat96}
C. Barrat and J.P. Garandet,  
Int. J. Heat Mass Transfer {\bf 39}, 2177 (1996).
%
\bibitem{Mey02} 
A. Meyer, Phys. Rev. B {\bf 66}, 134205 (2002).
%
\bibitem{MaMK04} 
S. Mavila Chathoth, A. Meyer, M. M. Koza, and F. Yuranji, 
Appl. Phys. Lett. {\bf 85}, 4881 (2004).
%
\bibitem{binder_book} 
D. P. Landau and K. Binder,
{\it A Guide to Monte Carlo Simulations in Statistical Physics}
(Cambridge University Press, Cambridge, 2000).
%
\bibitem{voter78}
A.F. Voter and S.P. Chen, in {\it Characterization of Defects in Materials},
edited by R.W. Siegel {\it et al.}, MRS Symposia Proceedings No. 82
(Materials Research Society, Pittsburgh, 1978), p.~175.
%
\bibitem{foiles87}
S.M. Foiles and M.S. Daw,
J. Mat. Res. {\bf 2}, 5 (1987).
%
\bibitem{das07}
S.K. Das, J. Horbach, and K. Binder, unpublished.
%

\endreferences
\end{document}